\documentclass[a4paper, 11pt]{article}

\usepackage[dvips]{graphicx}
\usepackage[usenames]{color}
\usepackage{amsmath}
\usepackage{mathtools}
\usepackage{changebar}
\usepackage{hyperref}

\newenvironment{revAdded}{}{}

\newcommand{\revDeleted}[1]{}

\begin{document}
\title{Digital Currency Design for Sustainable Active Debris Removal in Space\footnote{
This is a preprint (accepted manuscript) of the publication in
IEEE Transactions on Computational Social Systems (Volume: 6 , Issue: 1 ,
Feb. 2019), DOI: 10.1109/TCSS.2018.2890655
}}

\author{Kenji Saito\footnote{Keio Research Institute at SFC, Keio University},
Shinji Hatta\footnote{MUSCAT Space Engineering CO., Ltd.},
Toshiya Hanada\footnote{Department of Aeronautics and Astronautics,
Faculty of Engineering, Kyushu University}}
\date{}

\maketitle

\begin{abstract}
Orbital debris remains as an obstacle to further space development.
While efforts are ongoing to avoid newly launched objects becoming debris,
the number of debris would still continue to grow because of collisions.
ADR (Active Debris Removal) is an effective measure,
but building a sustainable economic model for ADR remains as a difficult
problem.

We propose that the cost of removal can be paid by circulating digital
currency tokens on a blockchain platform whose values may decrease and/or
increase over time, issued by global cooperation (a consortium) of parties
interested in space development, in exchange with proofs of ADR.
The tokens pay their cost by themselves through contributions by the token
holders, who are likely to be benefited by removal of debris.
This scheme imposes virtually no cost to the consortium.
We have generalized this concept as POD (Proof of Disposal), which, we believe,
provides a more accountable foundation for solving social problems with digital
currency than many ICO (Initial Coin
\begin{revAdded}or Cryptoasset\end{revAdded} Offering) in practice today.

We evaluated the feasibility of our proposal through a simulation.
We conclude that dynamic estimation of the economic values of each ADR and
automated pricing of tokens that represent the orbital debris being removed
are indeed possible.
Actual prototyping of the proposed digital currency system is ongoing.
\end{abstract}

\begin{description}
\item[Keywords:] Digital currency, Cryptocurrency, Blockchain,
Orbital debris,\\
Environmental remediation
\end{description}

\section{Introduction}

Orbital debris is a threat to active spacecraft and satellites.
Efforts are ongoing to ensure that newly launched objects will be
properly disposed from their orbits after their missions
(PMD: Post Mission Disposal).
However, the number of debris would still continue to grow even under ideal
conditions of no new launches, no debris release or no explosions, because of
collisions among existing orbital objects including debris themselves.
Therefore, for further development of space, it is mandatory for us to conduct
ADR (Active Debris Removal).
But building a sustainable economic model for ADR remains as a difficult
problem.

We see that a possible solution may be to design an economic medium, learning
from history.

Local currencies issued by local governments have been experimented in the
past, in order to make use of under-utilized resources in the region,
especially in times of depression.
Well-known cases include experiments in W\"{o}rgl, Austria, in 1932,
which made use of ``stamp scrip'' as a form of money whose value depreciates,
which helped the local government to invest on public works at virtually no
cost.
This design may be a hint for us to implement works of public interest with
limited budget.

The first author of this paper has extensively worked on possibilities of
implementing and utilizing such currencies as digital media that can be used
on the Internet.

\begin{figure}[h]
\begin{center}
\includegraphics[scale=0.35]{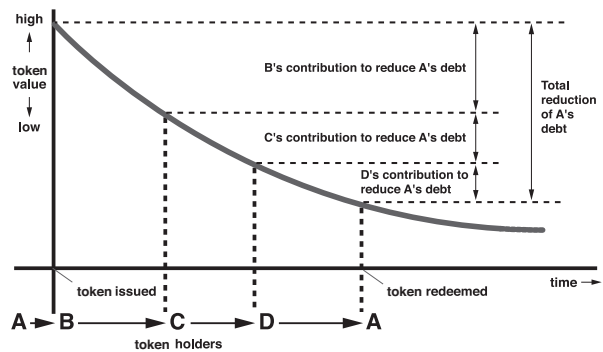}\\
\caption{Reduction-Over-Time and its Effects}\label{fig-rot}
\end{center}
\end{figure}
Figure~\ref{fig-rot} shows the effects of depreciating (Reduction-Over-Time)
currency tokens\cite{Saito2006:Iwat}.
While it helps to reduce the debt of the token issuer,
a game theoretic analysis showed that depreciating currency is likely to
accelerate people's spending.
\begin{figure}[h]
\begin{center}
\includegraphics[scale=0.35]{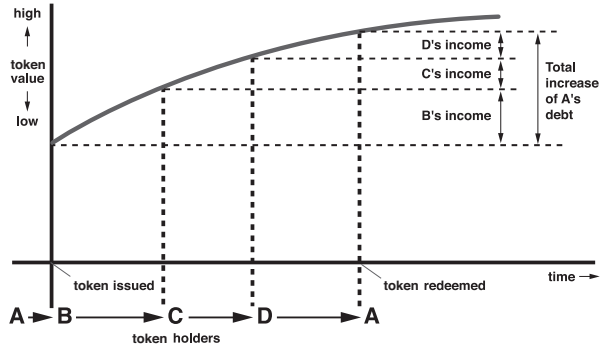}\\
\caption{Multiplication-Over-Time and its Effects}\label{fig-mot}
\end{center}
\end{figure}
Being digital, such media can easily be reversed to obtain opposite effects.
Figure~\ref{fig-mot} shows the effects of amplifying (Multiplication-Over-Time)
currency tokens\cite{Saito2005:MOT}.
An analysis showed that this type of currency decelerates people's spending.
Applications of such currencies have also been studied, for example,
for post-catastrophic disaster recovery\cite{Suko2007:PostCatastrophic}.

Unfortunately, these ideas have not received wide acceptance.
But situations are changing.

Bitcoin\cite{Nakamoto2008:Bitcoin} is now well-known and an accepted digital
currency.
Another example is Ethereum\cite{Buterin2014:Ethereum}, under development and
experimented widely, which make use of ``blockchain'', the record-keeping
foundation first developed for Bitcoin, as a platform of executing ``smart
contracts'' as distributed applications.
A blockchain, or a distributed ledger, is like a ``promise-fixation device in
the air'' that keeps records of promises, which can withstand partial failures
and {\em churns}.
It can be used for implementing an unstoppable monetary system because money
is essentially a promise that its recipient can also use it as money.

This paper proposes to utilize a new digital currency with planned
depreciation to build a sustainable economic model for ADR.
Then we will generalize this concept as POD (Proof of Disposal), which,
we believe, provides a more accountable foundation for solving social problems
with digital currency than many ICO (Initial Coin
\begin{revAdded}or Cryptoasset\end{revAdded} Offering) in practice today.

Remaining of this paper is organized as follows.
Section~\ref{sec-adr} gives background information on orbital debris and ADR.
Section~\ref{sec-design} shows the design of a digital currency that promotes
ADR by imposing its cost to users of the monetary tokens.
Section~\ref{sec-eval} evaluates the feasibility of the proposed design of the
currency by simulating issuance of tokens based on the real estimations of
collisions among orbital objects.
Section~\ref{sec-discuss} discusses potential societal influences and other
applications of the same design.
Finally, section~\ref{sec-conclusions} gives conclusive remarks.

\section{Orbital Debris and ADR}\label{sec-adr}

The instability of the orbital debris (OD) population in low Earth orbit
(LEO, the region below 2000km altitude), the ``Kessler Syndrome'', was
predicted by Kessler and Cour-Palais more than 30 years
ago\cite{Kessler1978:Debris}.
Recent modeling studies of the OD population in LEO suggested that the
current environment had already reached the level of instability.
Mitigation measures commonly adopted by the international space community,
including those of the Inter-Agency Space Debris Coordination Committee (IADC)
and the United Nations (UN), may be insufficient to stop the future population
growth.
In response to this new finding, an official IADC modeling study was conducted
in 2008 to assess the stability of the current environment.
Study participants were Agenzia Spaziale Italiana (ASI), British National
Space Centre (BNSC, now UK Space Agency, UKSA), European Space Agency (ESA),
Japan Aerospace Exploration Agency (JAXA) and National Aeronautics and Space
Administration (NASA).
The goal was to investigate the stability of the current environment using
the 1 January 2006 population as the initial condition.
The 200-year future projection adopted a ``best case'' scenario where no new
launches and no explosion beyond 1 January 2006 were allowed.
At the conclusion, a follow-up study, based on an updated environment
(including fragments from Fengyun-1C, Cosmos 2251, and Iridium 33), a more
realistic future lunch traffic cycle, and post-mission disposal 
implementation, was recommended.
Finally, the IADC designated the follow-up study as an official Action Item
(AI) 27.1, because of its potential significance.  The objective of AI 27.1
was to investigate the stability of the future environment and reach a
consensus on the need to use active debris removal (ADR) to stabilize the
future environment.
Participants included ASI, ESA, Indian Space Research Organization (ISRO),
JAXA, NASA, and UKSA.
Details of AI 27.1, its outcomes, and recommendations are summarized in
\cite{Liou2013:FutureLEO}.

In order to constrain the many degrees of freedom within IA 27.1, some
reasonable assumptions were made.  First, it was assumed that future launch
traffic could be represented by the repetition of the 2001 to 2009 traffic
cycle.  Second, the commonly-adopted mitigation measures were assumed to be
well-implemented.
In particular, a compliance of 90\% with the post-mission disposal ``25-year''
rule for payloads (i.e., spacecraft, S/C) and upper stages (i.e., rocket
bodies, R/Bs) and a complete passivation (i.e., no future explosions) were
also assumed.
However, collision avoidance maneuvers were not allowed, as in the previous
study.
In addition, an 8-year operational lifetime for payloads launched after
1 May 2009 was uniformly adopted.

Each participating member agency was asked to use its official, or best,
models for solar flux prediction, orbit propagation, and collision probability
calculation for AI 27.1.
Collision probability calculations were limited to 10cm and larger objects.
The NASA Standard Breakup Model\cite{Johnson2001:Breakup} was used by all
participants for their future projections, as it was determined that
participants did not employ any other fragmentation model.
The participants were encouraged to conduct as many Monte Carlo (MC)
simulations as time and resources allowed to achieve better statistical
results.
Finally, IA 27.1 conclusions were drawn primarily from the average results of
each participating model, determined through MC simulations.

\begin{figure}[h]
\begin{center}
\includegraphics[scale=0.33]{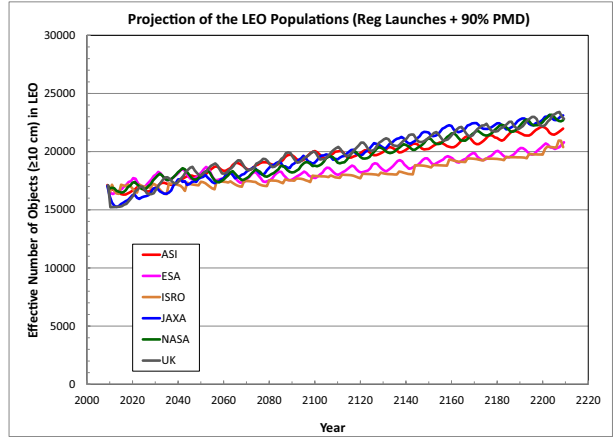}\\
\caption{Effective numbers of objects 10cm and larger in LEO predicted by
the six different models.
All models assumed no future explosion and 90\% compliance of the commonly
adopted mitigation measures.}\label{fig-leo-population-projection}
\end{center}
\end{figure}

Figure~\ref{fig-leo-population-projection} summarizes the projections of the OD
population in LEO through the
year 2209, assuming no future explosion and a 90\% compliance of the commonly
adopted mitigation measures, from the six models.
All models predict a future population growth.
The average increase is 30\% in 200 years.
The short-term fluctuation, occurring on a timescale of approximately 11 years,
is due to the solar flux cycle.

\begin{figure}[h]
\begin{center}
\includegraphics[scale=0.33]{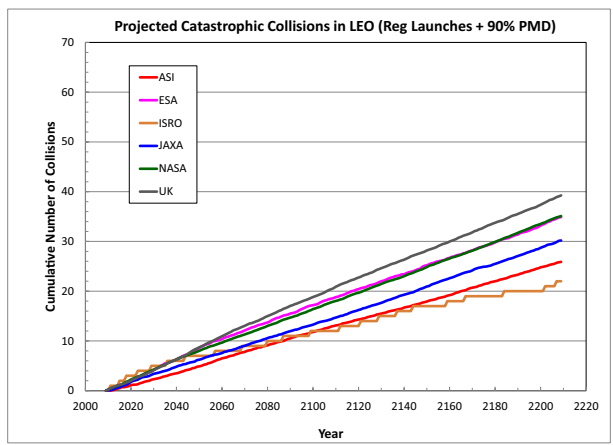}\\
\caption{Cumulative numbers of catastrophic collisions predicted by the six
models}\label{fig-leo-collisions-projection}
\end{center}
\end{figure}

Figure~\ref{fig-leo-collisions-projection} summarizes the cumulative number of
catastrophic collisions happened within the 200-year projection period.
Catastrophic collisions, such as the one between Iridium 33 and Cosmos 2251
in 2009, result in the complete fragmentation of the objects involved and
generate a significant amount of debris.
They are the main driver for future population growth.
The steepest curve (UKSA) represents a catastrophic collision frequency of
one event every 5 years, whereas the shallowest curve (ISRO) represents a
frequency of one event every 9 years.
Catastrophic collisions happened primarily at altitudes of 700-800km,
900-1000km.

The outcomes of AI 27.1 confirm the instability of the current OD population
in LEO.
They also highlight two key elements for the long-term sustainability of
outer space activities.
First, compliance of the mitigation measures, such as the 25-year rule,
is the first defense against the future population growth.
The need for a full compliance must be emphasized.
The 90\%-compliance assumption made in the simulations is certainly higher
than the current reality.
If the international space community cannot reach this level soon, future
population growth will be far worse than the outcomes of AI 27.1, and it
will certainly make future environment stabilization much more difficult.
Second, to stabilize the future environment, more aggressive measures, such as
ADR, must be considered.
Remediation of the environment after more than 50 years of space activities is
complex, difficult, and will likely require a tremendous amount of resources
and international cooperation.
The international community should initiate an effort to investigate the
benefits of environmental remediation, explore various options, and support
the development of the most cost-effective technologies in preparation for
actions to better preserve outer space for future generations.

\section{Design of Digital Currency for ADR}\label{sec-design}

\subsection{Overview of the Design}
We propose that the cost of ADR can be paid by circulating digital
currency tokens that depreciate over time, issued by global cooperation
(a consortium) of parties interested in space development, in exchange with
proofs of ADR.
We call the body of cooperation {\em the consortium} hereafter.
We call a party that conducts ADR {\em a remover}.
We call the monetary system we propose {\em the ADR currency}.

\begin{figure}[h]
\begin{center}
\includegraphics[scale=0.35]{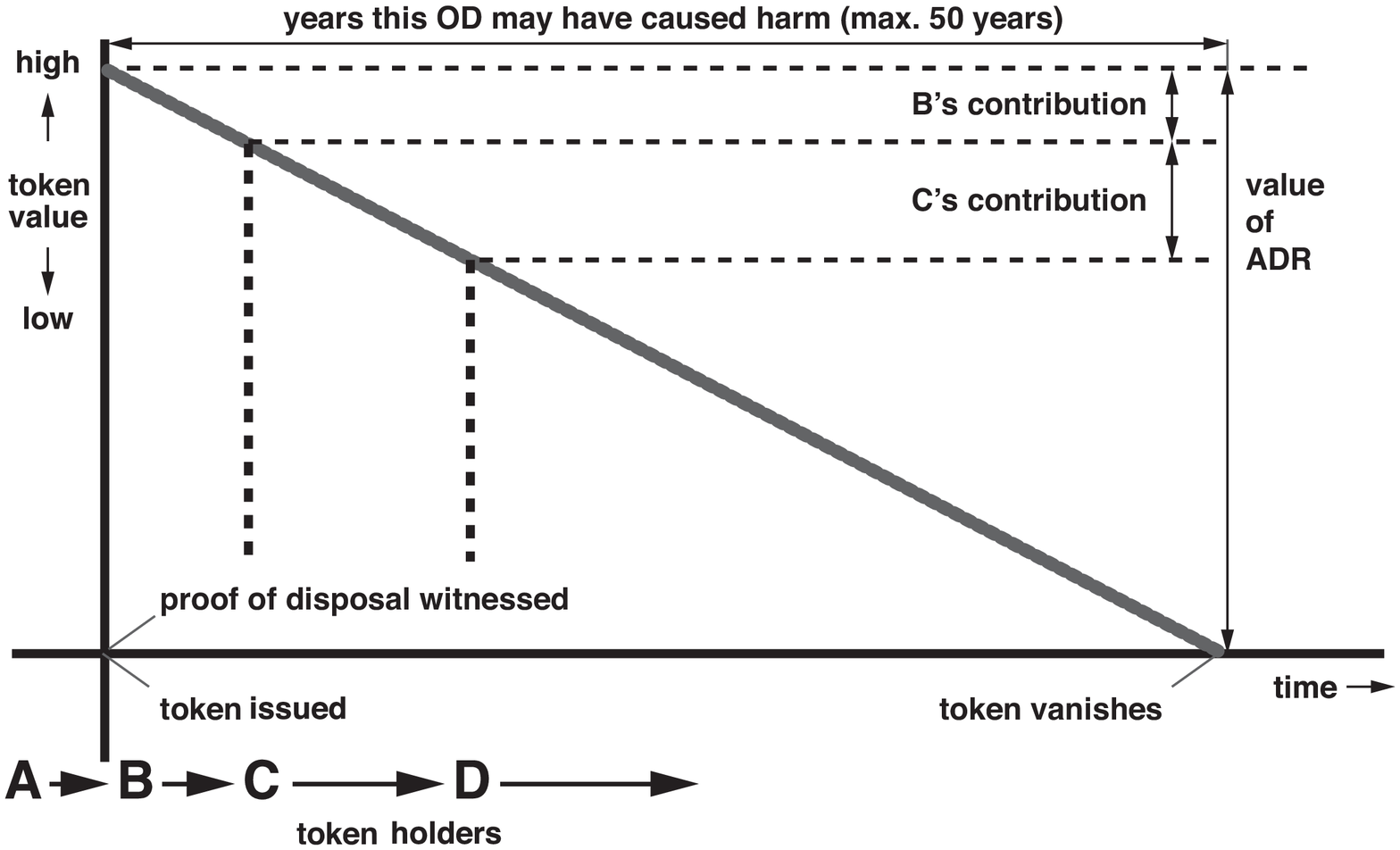}\\
{\footnotesize
\begin{itemize}
\item The issuer A is the consortium.
The first token holder B is a remover.
\item Depreciation in reality can be stepwise, such as yearly or monthly.
\end{itemize}
}
\caption{The ADR Currency}\label{fig-adr-currency}
\end{center}
\end{figure}

Figure~\ref{fig-adr-currency} shows the overview of the design of the
ADR currency.

\subsection{Initial Token Value}

Since the catalogue of all observed OD is available, we can calculate the
risk of collisions.
The consortium defines how the risk is calculated, representing the
international space community.
Probability of accidental collisions is calculated according to the defined
function, and then the collision flux is translated into a monetary value.
All calculations are conducted and published on the same blockchain platform
on which the ADR currency circulates, in order for all stakeholders to be
able to verify that these are correctly done.

In the translation, for example, the total cost of space development for the
past 60 years can be used.
The monetary value can be divided proportionally to the probabilities of
collisions, which is iteratively re-evaluated as ADR proceeds.

The initial token value is decided by the consortium according to the following
steps:
\begin{enumerate}
\item For each OD, the consortium periodically publishes the probability of
collisions and the estimated token value with its depreciation schedule
(the estimation expires after a certain duration of time).
\item A remover conducts an ADR (for the time being, a complete removal is
assumed for the sake of discussion), to which the consortium issues a token.
\item For each OD, the consortium recalculates the probability of collisions.
\end{enumerate}

\subsection{Depreciation}

An ADR currency token depreciates as Figure~\ref{fig-adr-currency} shows.

We would like to accelerate removal and decelerate the space development while
the number of debris is growing.
Originally upon conceiving the idea of the ADR currency, we thought that
the value of the tokens may be set to increase at first, to the extent allowed
by the consortium, and when the number of debris stops growing, the token
starts depreciating, thereby accelerating the industry.

However, the span for achieving the non-growth appears to be too long, such as
100$\sim$200 years.
Therefore, we have abandoned the idea of accelerating removal and decelerating
spending by increasing the token values for the ADR currency (this idea is
illustrated in section~\ref{sec-other-applications} for another application).

Instead, the tokens shall depreciate down to zero\footnote{If we decide that
the token value does not go down to zero, it will have to be redeemed by the
consortium when the minimum value is reached.
When this happens, the consortium will have to pay the value to the final
holder of the token.}
value over the duration of time the OD is estimated to cause harm, limited by,
for example, 50 years.

\subsection{Incentive Compatibility}

For a given OD, the higher its probability of collisions is, the higher the
initial token value would be when the ADR is performed.
The longer the duration of time the OD would cause harm, the more slowly the
corresponding token would depreciate.
As a result, a remover would want to aim for removing an OD with higher
probability of collusions that would stay for a longer time.
The incentive provided by the currency design is compatible with our
intention to make future space development safer and easier.

The tokens pay their cost by themselves through contributions by the token
holders, who are likely to be benefited by the removal of the OD.
This scheme imposes virtually no cost to the consortium.

\section{Feasibility}\label{sec-eval}

Evaluation of the effects of the proposed currency design will involve how the
industry and market react to this design, which is not objectively
estimated\footnote{Effects of depreciating/amplifying currencies have been
evaluated in depth by simulations with some assumed human behaviors in
\cite{Saito:2010:BSR:1838759.1838822}, for example.}.
Instead, we evaluate the feasibility of the currency system by simulating
issuance of tokens using the actual catalogue data and orbital projections.

For the matter of discussion, we chose 3,866 intact objects in LEO, i.e.
uncrushed objects such as rocket upper fuselage or satellite platforms, from
the catalogue data as of April 1, 2017.
The objects include operational satellites, as we cannot distinguish between
unused objects and ones in use, unless the intentions are confirmed.
 
We divide space into $\frac{10}{3}$-kilometer cubes, and locate each object in
a cube at a given time.
Then we examine whether each object has any neighbors in the cube they are
located or in any of the adjacent cubes
(this means that we consider a 10-kilometer cube for each object).
If such a neighbor is found, the collision flux is calculated according to
the relative velocity of the objects.
We iterate this procedure with one-day interval for 50 years in our simulated
time to estimate the accumulated collision flux, or risk of collisions,
for each object.

\begin{figure}[h]
\begin{center}
\includegraphics[scale=0.45]{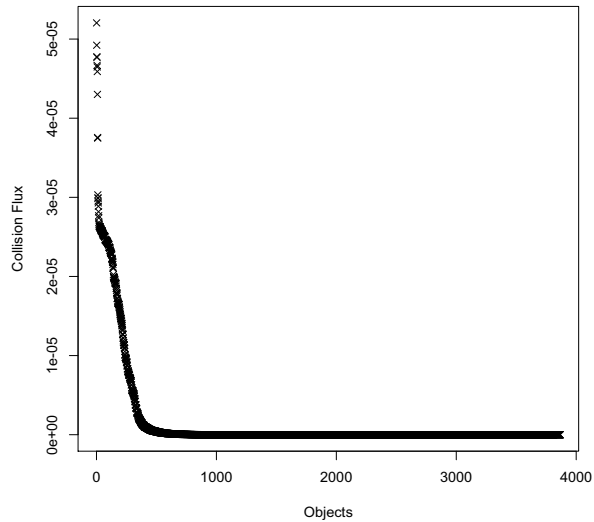}\\
\caption{Initial Distribution of Accumulated Collision Flux}
\label{fig-collision-flux}
\end{center}
\end{figure}

Figure~\ref{fig-collision-flux} shows the initial distribution of accumulated
collision flux.

We consider series of ADRs where an object with the highest accumulated
collision flux is removed from the orbit at a time.
We approximate the effects of an ADR by removing the corresponding item from
the same catalogue data and recalculating the fluxes.

\begin{figure}[h]
\begin{center}
\includegraphics[scale=0.45]{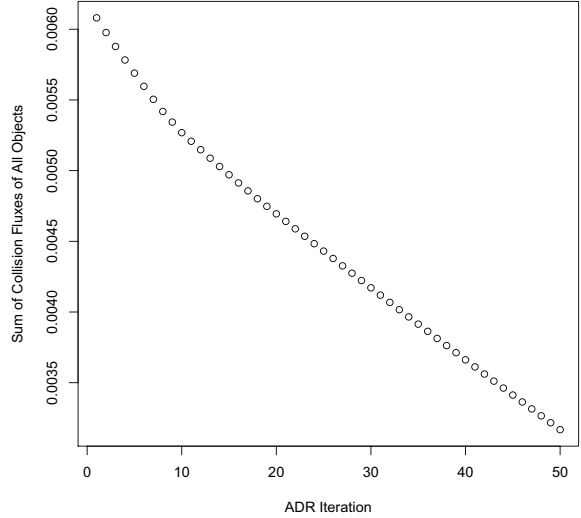}\\
\caption{Total Collision Fluxes}
\label{fig-total-collision-fluxes}
\end{center}
\end{figure}

\begin{figure}[h]
\begin{center}
\includegraphics[scale=0.45]{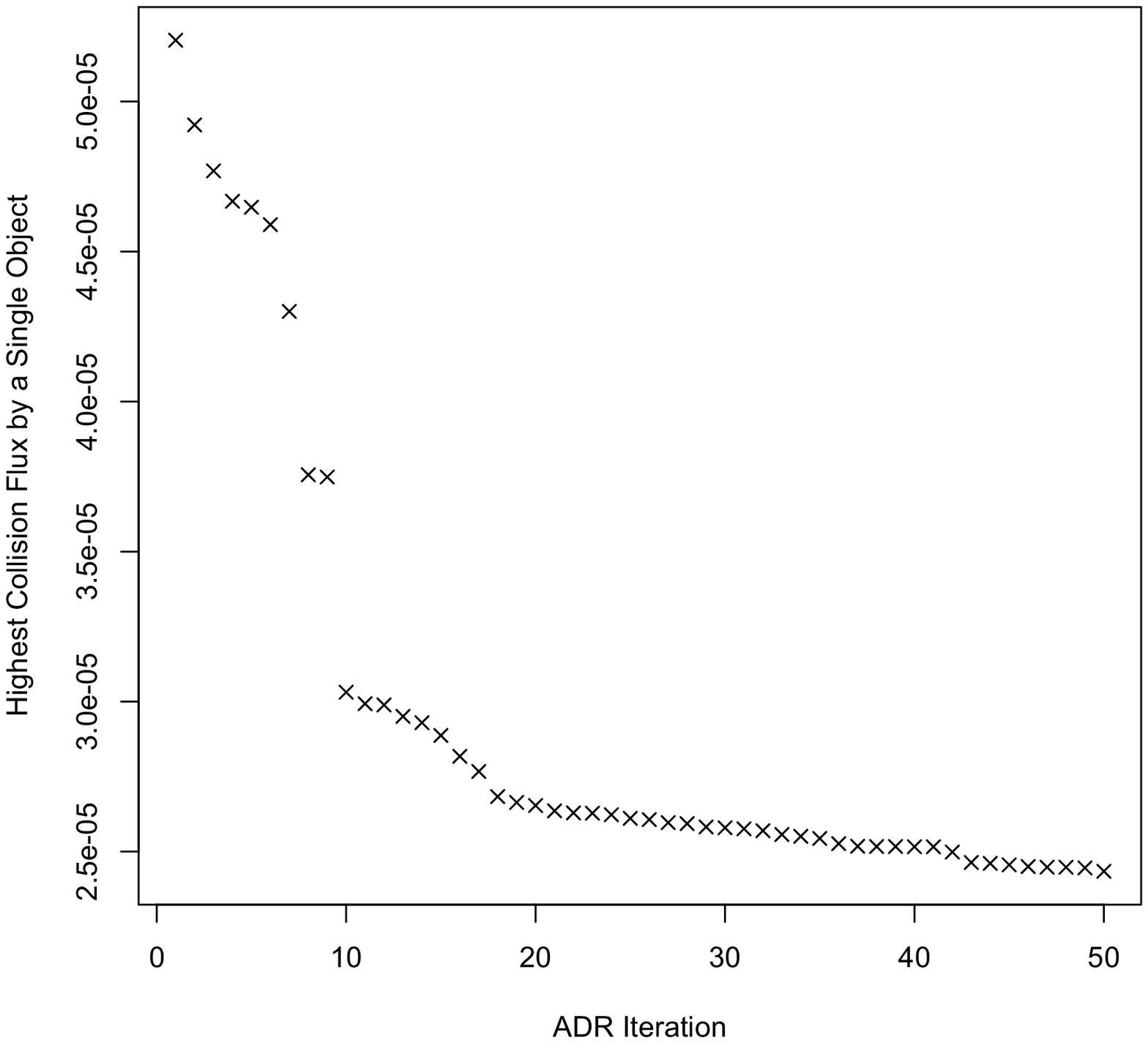}\\
\caption{Single Object Collision Fluxes}
\label{fig-single-collision-fluxes}
\end{center}
\end{figure}

Figure~\ref{fig-total-collision-fluxes} shows changes of the sum of collision
fluxes of all objects, and 
Figure~\ref{fig-single-collision-fluxes} shows 
the highest collision flux by a single object (i.e. the flux
of the removed object), over 50 iterations of ADRs as described above.
We see monotonous decreases in both values.

For each orbital object, we assign an initial token value proportionally to
its accumulated collision flux.
To do so, we set the initial virtual budget of 3.2 trillion USD, referring to
our estimation of the total cost of space development for the past 60 years,
which is the debt the space industry and the surrounding communities should be
willing to repay over many years by taking the depreciating tokens as payments.
For the first iteration of ADRs, we divide the budget proportionally to the
accumulated collision flux of each object, and assign the value as the initial
value of the token representing the object.

We consider two different policies for pricing for later iterations.

\begin{description}
\item[division policy:] $\:$\\
We subtract the initial token value of the removed object from the budget, and
recalculate in the same way.
Over time, the sum of initial token values of removed objects will never
exceed the initial virtual budget.

\item[coefficient policy:] $\:$\\
We obtain the coefficient of translating the collision flux into a token value
from the first iteration, and use the coefficient for later iterations.
Over time, the sum of initial token values of removed objects may exceed the
initial virtual budget.
\end{description}

\begin{figure}[h]
\begin{center}
\includegraphics[scale=0.4]{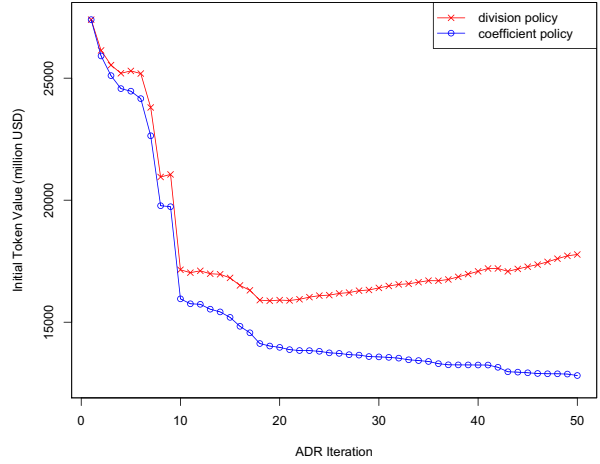}\\
\caption{Highest Initial Token Values}\label{fig-initial-token-values}
\end{center}
\end{figure}

Figure~\ref{fig-initial-token-values} shows the highest initial token values
(i.e. the token values of removed objects) for the first 50 iterations of ADRs,
calculated in these two different policies.
Note that in this work, these tokens are valued for demonstration and
prototyping purposes only,
and the resulted token values do not necessarily suggest how much each ADR
should be valued in reality.

With {\em division policy}, we see that the highest initial token value may
increase as contribution rate of the collision flux of the removed object
against the whole may increase over time.
With {\em coefficient policy}, on the other hand, the price monotonously
decreases, as long as the highest collision flux keeps decreasing.
For incentive reasons, monotonous decrease of highest initial token values
is preferred, because it would not motivate removers to wait.

However, we should note that this simulation assumes no collisions happening
(thus assuming no increase of objects), and in reality, the highest collision
flux would not keep decreasing even with some ADRs.
We may need more sophisticated policy to realize the monotonous decrease of
prices.

\section{Discussion}\label{sec-discuss}

\subsection{``Proof of Disposal'' concept}

In addition to physical or engineering aspect of OD including
intact object described in section~\ref{sec-adr}, the social aspect of OD
and the proposed scheme is considered in this section, for generalization of
the scheme for other applications.

\begin{figure}[h]
\begin{center}
\includegraphics[scale=0.5]{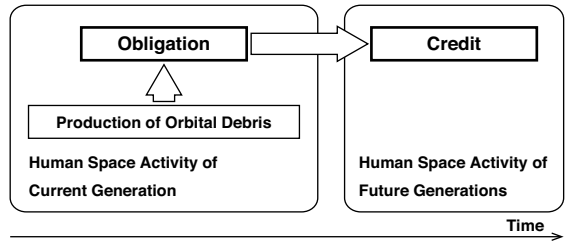}\\
\caption{Obligation and Credit}\label{fig-obligation}
\end{center}
\end{figure}

We try to understand the proposed currency system in terms of obligation and
credit.
Once an OD is produced, the risk diffuses thinly and broadly to all orbits
that share the same space with the object.
The responsible body for the object does not suffer any disadvantage except for
the thinly diffused risk.
To make matters worse, the risk diffuses not only spatially but also
temporally.
The temporal spread of the risk is, in other words, a negative legacy from
human space activity of the current generation to that of the future
generations.
Every time an OD is produced, human space
activity of the current generation increases corresponding obligation to that
of the future generations, and in consequence, human space activity of the
future generations increases credit to that of the current generation
(Figure~\ref{fig-obligation}).

\begin{figure}[h]
\begin{center}
\includegraphics[scale=0.5]{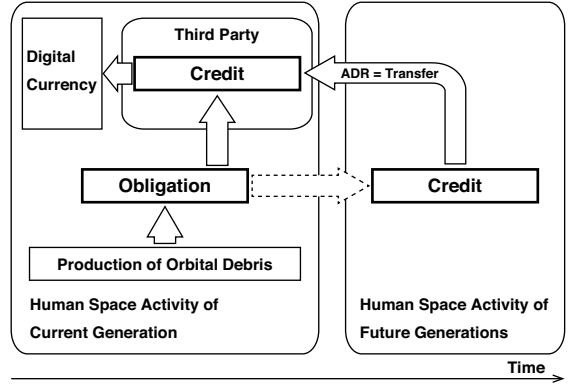}\\
\caption{Transfer of Credit from Future to Today}\label{fig-credit-transfer}
\end{center}
\end{figure}

It seems impossible for a creditor in the future to recover the credit from
the current obligor.
It is possible, however, for some third party to transfer the credit from the
future creditor through a present-day activity, namely ADR.
The third party to conduct ADR reduces the negative legacy through the
environment remediation, and therefore, ADR is considered equivalent to
transferring the credit from the future to today.
The scheme is shown in Figure~\ref{fig-credit-transfer}.

The transferred credit is a warranty for issuing a claim deed to be
circulated as a digital currency token.
The authors have named this scheme for issuing digital currency
{\em Proof Of Disposal}, or POD in short.

The scheme may be applicable for any kind of disposal of industrial or war
wastes.
For establishment of POD, however, the corresponding disposal has to be
performed under the witness of the whole community, in order to prevent an
untrue statement of disposal, and the knowledge of the disposal is presumably
to be shared among the community members through the same blockchain platform
that runs the currency system.
From this point of view, ADR is one of the most suitable cases for POD.

\begin{revAdded}
\subsection{Practical Values of POD Tokens}

But would POD (in particular, ADR currency) tokens have any practical values,
if they are issued by consortia of parties instead of banks, and with
scheduled depreciation?

Because ADRs remediate space environment for a wide range of applications
including weather forecasting, maritime, broadcasting, telecommunication,
airline and navigation,
the consortium would actually involve many companies not directly associated
with space development.
In return, the currency system would attract lots of customers of such
companies, who would want to use the tokens to pay for the services (the
currency accelerates people's spending by design), such as mobile
communication fees, for example.
There may be a large body of the users who find practical values in the tokens.

There is also a reason why POD tokens need to be linked with existing monetary
values.
If such tokens are independent from other currencies, depreciation would not
be effective for decreasing the token values as expected, because then tokens
become scarcer, and more highly valued because of less supplies against
demands in relative to other currencies.
Therefore, POD tokens need to be pegged to an existing currency, such as US
dollar.
A regulatory way to ensure such pegging may be to regard depreciation as
donation to the consortium (because depreciation does reduce the debt of the
consortium), which may be made tax-deductible in countries where the
consortium is considered a charitable organization.

\end{revAdded}

\subsection{Related Work}

\begin{revAdded}

\subsubsection{Economics of Orbital Debris}
Both
\cite{Adilov2015:Pigovian} and
\cite{Salter2016:Debris} provide economic analyses of the problem of OD.
The former proposes that classic Pigovian taxes on new launches would reduce
creation of new debris, and could also possibly fund ADR.
But it would require cooperation among competing nations that have created
the majority of extant debris, and spontaneous deceleration of new launches
would not incentivize these nations.
The latter argues that the most reliable solution to the kind of problem is
the establishment and enforcement of private property rights, which they find
infeasible in this case.
It also suggests that an institution to work on this problem would be
classified along a spectrum that transcends the private-public dichotomy.

Indeed, our proposal of POD is based on our understanding that the solution
would not come from the dichotomy but from the civil society.

\end{revAdded}

\subsubsection{Funding and Internet Technology}

In recent years, a way of funding from general public through the Internet
has become a popular practice.
It is called
{\em crowdfunding}\begin{revAdded}\cite{Riedl2016:Crowdfunding}\end{revAdded}.
However, crowdfunding is not necessarily a solution for implementing public
works where sufficient money is unavailable, as it requires money to be
gathered first.
In the proposed scheme of POD, on the other hand, monetary medium is created
on demand on a blockchain platform.

Another way of funding that is becoming a popular practice is {\em ICO (Initial
Coin \begin{revAdded}or Cryptoasset\end{revAdded}
Offering)}\begin{revAdded}\cite{Hartmann2018:ICO}\end{revAdded}, or
a {\em crowdsale} of newly issued digital coins on a
blockchain platform.
Although both ICO and POD issue digital currency tokens,
ICO is rather similar to crowdfunding, as it requires upfront money to
purchase the coins.
ICOs tend to attract more people more rapidly than crowdfunding, as people
expect capital gains with an anticipation that the price of the coins will go
up in the future.
Because of this expectation, circulation as monetary medium is limited for
the issued coins in ICO (spending is generally decelerated).
Thus, the issued digital coins are not expected to help the community as a
whole as means for payment among community members.
Another point about ICO is that issued digital coins do not usually represent
debt, so that ICO requires an external scheme for monitoring whether the
proposed project is soundly ongoing or not.

\begin{revAdded}
\subsubsection{Incentives and Attractions for Environmental Remediation}

Plastic Bank\cite{PlasticBank:Web} is another example of turning waste into
a monetary medium,
where recycling plastic is rewarded with
digital tokens issued on some blockchain platform, in order to stop ocean
plastic and help people ascend from poverty.
Plastic Bank incentivizes collectors of ocean plastics by 
paying them a premium on top of the market rate, using funds raised by
crowdfunding.

This idea is similar to POD.
But with POD, we can unnecessitate crowdfunding, and distribute the cost of the
premium among receivers of the issued tokens, while spreading and promoting the
idea along with circulation of the tokens at the same time.

However,
whether we use crowdfunding or POD, an important question would be how helping
environmental remediation can attract people and change their behaviors.
\cite{Tushar2018:AC}, for example, proposes a design inspired by motivational
psychology, towards more environment-friendly use of air conditioning.
\cite{Hu2018:Incentive} studies robustness of incentive mechanisms with
bounded rational behaviors.
\cite{Sajadi2018:SocialNorms} studies the spread of social norms.
We can learn from those studies how to attract people and organize people's
participation in our proposed solution to the problem of OD.

\end{revAdded}

\subsection{Other Applications}\label{sec-other-applications}

\begin{figure}[h]
\begin{center}
\includegraphics[scale=0.35]{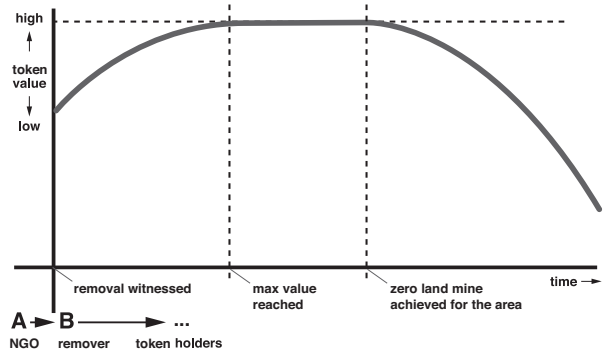}\\
\caption{Zero Land Mine Currency}\label{fig-zerolandmine}
\end{center}
\end{figure}

POD is expected to work where disposal is observed publicly.
Another possibly useful case is land mine clearing
(Figure~\ref{fig-zerolandmine}).
This example uses both amplifying and depreciating features of digital currency
tokens issued in return for disposal.
First, the issued tokens amplify their values for both incentivizing early
removal of land mines and decelerating spending.
Depreciation is delayed until zero land mine is achieved in the defined
area, after which the area can be reconstructed with accelerated spending.

\subsection{Implementation and Prototyping}

We believe that either Bitcoin or Ethereum can be used for implementing POD.
However, both blockchain platforms are under a risk of being discontinued.
If the price of the native token (bitcoin or Ether, respectively) goes down,
the maintainers of the blockchain (often called {\em miners}) may have to
leave the system, because the cost of maintaining the blockchain, which is
usually covered by rewards they receive in the form of native tokens, would
be valued more than the rewards.

To ensure sustainability of the ADR currency, which would have to continue for
many tens of years, we take part in developing BBc-1 (Beyond Blockchain
One)\footnote{
https://github.com/beyond-blockchain/bbc1}\begin{revAdded}\cite{Saito2017:BBc1}\end{revAdded},
a new open-source platform for record-keeping that can use either Bitcoin or
Ethereum (and even dynamically switch between them) for proof of existence of
transactions in its early stage (and later, it will become independent).
We are prototyping the ADR currency on top of this new platform.

\section{Conclusions}\label{sec-conclusions}

We proposed that the cost of removal of OD can be paid by circulating digital
currency tokens whose values decrease over time, issued in exchange with
proofs of ADR.
We have shown that managing such a currency system, the ADR currency, is
indeed feasible.

We consider that OD is a kind of debt owed by the human race of existing
generation to the future generations.
The ADR currency has an effect of transferring the corresponding credit to
the present-day.
This concept can be generalized for many social debts publicly shared, for
starting and managing projects for social goods using blockchain platforms.

This is also an implementation of a belief that for a currency to be
accountable, creation of monetary amount
should be warranted by some kind of value in the real world, as opposed by
many cases of ICO.
The authors hope that the scheme we proposed in this work will be considered
as an alternative to crowdfunding or ICO for implementing public work as
blockchain applications under low budget conditions.

\section*{Acknowledgements}

We express our sincere thanks to Prof.~Setsuko Aoki, the vice director of
Center for Space Law, Keio University, for her advice from the viewpoint of
space law.

\bibliographystyle{plain}
\bibliography{adr-currency}

\end{document}